\documentclass[]{aastex701}
\usepackage{savesym}
\usepackage{amsmath}
\savesymbol{iiint}
\usepackage{wasysym}

\definecolor{DarkGreen}{rgb}{0.0,0.4,0.0}  
\newcommand{\B}[1]{{\color{blue}{#1}}}
\newcommand{\R}[1]{{\color{red}{#1}}}
\newcommand{\G}[1]{{\color{DarkGreen}{#1}}}
\newcommand{\M}[1]{{\color{magenta}{#1}}}
\newcommand{\C}[1]{{\color{cyan}{#1}}}
\usepackage[normalem]{ulem}
\usepackage{soul}
\usepackage{cancel}

\begin{document}

\title{Height and Energy Evolution of X-ray Double Sources in a Solar Flare}

\author[orcid=0000-0002-0318-8251,sname='Pan']{Hanya Pan}
\affiliation{CAS Key Laboratory of Geospace Environment, Department of Geophysics and Planetary Sciences, University of Science and Technology of China, Hefei 230026, China}
\affiliation{Institute of Physics, University of Graz, A-8010 Graz,Austria}
\affiliation{Zhejiang Wenzhou High School, Wenzhou 325014, China}
\email{hanyapan@mail.ustc.edu.cn}  

\author[orcid=0000-0003-2073-002X,sname='Veronig']{Astrid M. Veronig} 
\affiliation{Institute of Physics, University of Graz, A-8010 Graz,Austria}
\affiliation{Kanzelhöhe Observatory for Solar and Environmental Research, University of Graz, A-9521 Treffen, Austria}
\email[show]{astrid.veronig@uni-graz.at}

\author[orcid=0000-0003-4618-4979,sname='Liu']{Rui Liu} 
\affiliation{CAS Key Laboratory of Geospace Environment, Department of Geophysics and Planetary Sciences, University of Science and Technology of China, Hefei 230026, China}
\affiliation{CAS Center for Excellence in Comparative Planetology, Hefei 230026, China}
\email[show]{rliu@ustc.edu.cn}

\correspondingauthor{Astrid M. Veronig and Rui Liu}

\begin{abstract}

In the standard model, magnetic reconnection at a vertical current sheet above the flare arcade is key to explaining many aspects of solar eruptions.
The supra-arcade region is where the vertical current sheet is supposedly located, and X-ray/EUV emission therein reflects underlying energy release and transport processes, therefore providing valuable insight into the structure and evolution of the current sheet. Previous studies have focused primarily on the impulsive phase of flares, but phenomena in the decay phase are also crucial for understanding the complete flaring scenario.  
In this paper, we investigated an M6.7-class limb flare that occurred on August 28, 2022, combining observations from the Solar Orbiter (SolO) and Solar Dynamics Observatory (SDO). Coronal X-ray sources are continually observed by the Spectrometer/Telescope for Imaging X-rays (STIX) onboard SolO for over two hours, revealing a multi-phase evolution with varying velocities and multiple substructures, with higher-energy components consistently appearing at higher altitudes.
Such a height-energy relation is notably observed in a double coronal source during the decay-phase, which is dominated by thermal emission. The energy distribution of the double source distinguish itself from previous studies that showed a symmetric distribution, with the higher-energy components being closer to the center of the double source during the impulsive phase. Obtained from two opposite side-on perspectives on the supra-arcade region, these findings reveal the spatio-temporal complexity of the energy release process in the post-flare current sheet during the decay phase.

\end{abstract} 



\section{Introduction} \label{sec:intro}
In the standard eruptive flare model, the current sheet beneath the erupting magnetic structure is the key region where magnetic reconnection converts magnetic energy into plasma heating and particle acceleration \citep{Carmichael1964, Sturrock1966, Hirayama1974, Kopp1976}.  
In a two-dimensional (2D) scenario, the current sheet forms between oppositely directed magnetic field lines that are stretched by the eruptive structure \citep{Kopp1976, Liu2013}. 
In a three-dimensional (3D) scenario, it forms along hyperbolic flux tubes (HFTs) or quasi-separatrix layers (QSLs), where magnetic connectivity changes rapidly \citep{Priest1995, Janvier2014}.  
As the eruptive magnetic structure ascends, the current sheet becomes progressively longer and thinner.  
It can persist for several hours after the initial ejection, continuing to extend both in length and to rise in height \citep{Patsourakos2011,Lin2015}.

Observationally, an elongated plasma sheet is considered as a proxy for the current sheet when viewed edge-on, formed by heated plasma surrounding the thin current sheet itself \citep{Ko2003, Webb2003, Lin2005}.
When viewed side-on, the region where the current sheet is expected often appears as a diffuse, heated plasma located above the post-flare loops, known as the supra-arcade fan \citep[SAF;][]{Svestka1998}.
This structure is typically observed in X-ray and EUV images, with temperatures around 10 MK \citep{McKenzie1999, Hanneman2014, Innes2014, Cai2019, Savage2012}, and is more commonly seen in long-duration flares \citep{McKenzie2001}.
\citet{Svestka1998} first identified the SAF using Yohkoh/SXT observations and proposed that it represents plasma flowing along open magnetic field lines in the associated active region (AR), eventually contributing to the solar wind.
An alternative interpretation suggests that the SAF may involve ongoing magnetic reconnection, which is supported by the spatial coincidence of SAF and the expected current sheet. 
Moreover, violent lateral motions have occasionally been detected within the fan region \citep{McKenzie1999}, associated with dark downflows observed in soft X-rays, known as supra-arcade downflows (SADs) \citep{McKenzie1999}.
SADs are thought to be evidence of reconnection-driven flows originating from high altitudes \citep{Forbes1996}.

X-ray emission is also a key observational signature predicted by the standard flare model. 
Following magnetic reconnection, energetic electrons are accelerated in the solar corona. When interacting with ions in the dense chromosphere, they produce hard X-rays through nonthermal bremsstrahlung radiation.  
During the impulsive phase of solar flares, hard X-ray sources typically appear at the footpoints of flare loops or ribbons, where energetic electrons precipitate and are stopped in the dense chromosphere \citep{Hoyng1981, Kane1983}.  
In addition, hard X-ray sources are also observed at loop tops, weaker than footpoint sources for most cases. They are attributed to interactions between accelerated electrons and the loop-top plasma \citep{Krucker2008}.
As mentioned above, X-ray emission in the supra-arcade region, which is closely associated with the vertical current sheet, are useful for the diagnostics of the energy-release processes in flares.

Occasionally, a second, fainter hard X-ray source appears above the loop-top source \citep{Sui2003,Sui2004,Liu2008}.
They are together referred to as a `coronal double source'. 
\citet{Sui2003} first reported a pair of X-ray sources: one located at the flare loop top and the other approximately 25 Mm above it.  
These sources exhibited a symmetric, energy-dependent height distribution, with higher-energy emissions concentrated near the center, suggesting the presence of an X-point between them.  
Such a double-source structure is reported in different events, and different formation mechanisms have been suggested \citep{Sui2004, Veronig2006, Li2007, Liu2008, Chen2012, Liu2009, Liu2013, Ning2016, Su2013, Kim2014}.
In many cases, the double source is believed to result from bidirectional outflows from a reconnection current sheet, which is aligned with the scenario proposed by \citet{Sui2003}. For instance, \citet{Liu2008} explained the symmetric double source during the impulsive phase of a flare using a stochastic acceleration model \citep{Hamilton1992, Miller1996}.  
The double source has also been linked to the X-point in a breakout reconnection geometry \citep{Su2013}, but the dominance of thermal emission leaves room for different interpretations \citep{Li2007, Kim2014, Kolomanski2017, Zhou2022}. For example, \citet{Kim2014} observed a double source where the upper source features increasing temperature and density during the main and decay phases, which is interpreted by an upward energy transport.

Most studies of X-ray sources focus on the impulsive phase, during which the energy release is most rapid. These observations have revealed several typical spatial evolution patterns.
One is the height variation, i.e., a pre-flare downward motion of the  coronal source, followed by an upward rise during the main phase, which has been attributed to different processes, like spectral hardening, magnetic relaxation, flux tube retraction, or collapsing trap effects \citep{Sui2004,OFlannagain2013,Ji2007, Shen2008,Veronig2006,Unverferth2023,Li2006}. 
Another pattern is the size evolution: the flare loop volume, length, and thickness shrink before the X-ray peak and expand afterward, suggesting contraction followed by expansion \cite[]{Jeffrey2013}. The broadening after the peak may reflect magnetic diffusion caused by turbulence, allowing cross-field electron transport.

Decay-phase observations are equally important for understanding the long-term evolution of the flare dynamics and of the associated CME \cite[]{Gou2020}.
After the main energy release, long-lived current sheets have been observed extending outward from the flare site, often accompanied by plasma blobs or SADs moving along them \citep{Ciaravella2002, Ko2003, Lin2005, Liu2010, Sheeley2002, Innes2003, Savage2010, Awasthi2022}.
\citet{Saint-Hilaire2009} reported a RHESSI X-ray source that persisted for 12 hours, initially located above the flare loop and rising slowly over time.
The source showed a slight upward shift of higher-energy X-rays relative to lower-energy ones, which was interpreted as hot thermal emission from the base of the current sheet beneath the erupting structure.
\citet{Gallagher2002} presented similar findings from RHESSI and TRACE\footnote{Transition Region and Coronal Explorer} observations of the 21 April 2002 X1.5 flare, where a coronal X-ray source rose to over 140 Mm over 12 hours. TRACE also revealed a fan-like system of dark sinuous lanes extending from the flare arcade during the decay phase, possibly associated with downward-propagating voids or plasma instabilities. 
Such decay-phase observations provide critical insight into the sustained magnetic reconnection and gradual restructuring of the corona after major eruptions.

In this paper, we use observations from SolO and SDO to study a limb flare on August 28, 2022, during which its loop-top X-ray source is continually tracked for over two hours. The paper is organized as follows:  \S\ref{sec:methods} describes the instruments we used and the methods we developed to estimate the X-ray source height; \S\ref{sec:obs} presents the observational results; and \S\ref{sec:conclusion} provides concluding remarks.

\section{Instruments and Methods} \label{sec:methods}
\subsection{Instruments}
Launched on 10 February 2020, SolO \citep{Muller2020} is the first mission of ESA's Cosmic Vision 2015--2025 program and a joint effort with NASA.  
It carries six remote-sensing instruments and four in-situ instruments to study the Sun.  
Its orbit includes multiple gravity assists from Venus and Earth, gradually bringing it closer to the Sun, with perihelia reaching as close as 0.28 AU.
Over time, the inclination of the spacecraft's orbital plane relative to the ecliptic will gradually increase, eventually exceeding 30$^{\circ}$ and enabling unique views of the Sun's polar regions.

The Spectrometer/Telescope for Imaging X-rays \citep[STIX;][]{Krucker2020} is a hard X-ray imaging spectrometer aboard SolO, providing data products for X-ray spectral analysis and imaging reconstruction. 
It uses passively cooled Cadmium Telluride (CdTe) detectors to perform spectral observations in the 4--150 keV range, with an energy resolution of 1 keV.
STIX employs an indirect Fourier imaging technique, using a pair of tungsten grids placed in front of 32 CdTe detectors. 
The grid spacing ranges from 0.038 to 1 mm, yielding an angular resolution between 7 and 180 arcseconds.
It offers an imaging field of view of 2$^{\circ} \times$2$^{\circ}$, a positional accuracy of 4 arcseconds, and a finest angular resolution of 7 arcseconds. 
Depending on the count statistics, the time resolution may reach sub-second scales.
The imaging method was used in Japan's Yohkoh mission \citep{Kosugi1991}, and later adopted and further developed in NASA's Reuven Ramaty High-Energy Solar Spectroscopic Imager \citep[RHESSI;][]{Lin2002}. More recently, China's Advanced Space-based Solar Observatory \citep[ASO-S;][]{Zhang2019,Gan2019} employs a similar technique, which draws upon the concept developed for STIX.

The core principle of STIX Fourier imaging is to utilize the angular offset between each pair of sub-collimators to produce large-scale Moir\'e patterns on the detectors from incoming parallel X-rays.
The amplitude and phase of these patterns encode the Fourier components (visibilities) of the X-ray source's angular distribution \citep{Krucker2020}.
The limited number of visibility measurements constrains the complexity of structures that can be reconstructed.  
Therefore, STIX primarily targets relatively strong sources, with a design goal of achieving a dynamic range of 20:1 \citep{Krucker2020}.
Reconstruction methods are generally classified into count-based (e.g., CLEAN, EM, Pixon) and visibility-based (e.g., MEM\_GE, VIS\_FWDFIT) approaches.
In this work, we primarily adopt MEM\_GE \citep{Massa2020}. For reference, we also considered results from CLEAN \citep{Hogbom1974} and EM \citep{Benvenuto2013,Massa2019}, which exhibit trends very similar to those of MEM\_GE. However, these results are not presented as figures in this paper.

The extreme-ultraviolet (EUV) data used in this study are obtained from the Atmospheric Imaging Assembly \citep[AIA;][]{Lemen2012} on board SDO \citep{Pesnell2012} and the Extreme Ultraviolet Imager \citep[EUI;][]{Rochus2020} on board SolO.  
AIA captures full-disk solar images with a 41$\times$41 arcminute field of view, at a size of 4096$\times$4096 pixels (0.6 arcseconds per pixel), achieving a spatial resolution of 1.5 arcseconds and a temporal cadence of 12 seconds.  
It uses filters covering ten different wavelength channels, corresponding to specific emission lines targeting on different solar atmospheric regions.  
In this study, we primarily use AIA data from the 94 \AA, 131 \AA, 171 \AA, and 1600 \AA~passbands.
EUI captures both full-disk and high-resolution images (localized regions).  
It consists of three telescopes: the Full Sun Imager (FSI) and two High Resolution Imagers (HRI).  
FSI observes the Sun in the 174 \AA~and 304 \AA~EUV channels with a field of view of 3.8$^{\circ} \times$ 3.8$^{\circ}$. Its images have a size of 3072$\times$3072 pixels, corresponding to an angular resolution of 10 arcseconds.
To further investigate the thermal properties of the plasma, we apply differential emission measure (DEM) analysis based on AIA's EUV data, following the methods developed by \citet{Cheung2015} and \citet{Su2018}.
Due to the relatively low signal-to-noise ratio in the EUV data within the region of interest at the time under study, we improved the data quality by averaging three sequential AIA images---namely, the target frame and its immediate temporal neighbors. The resulting averaged image served as the input for the subsequent DEM calculation.

The Helioseismic and Magnetic Imager \citep[HMI;][]{Scherrer2012} aboard the SDO satellite provides full-disk line-of-sight (LoS) magnetic field maps (4096$\times$4096 pixels) with a temporal cadence of 45 seconds, and a spatial resolution of 1 arcsec.
The Polarimetric and Helioseismic Imager \citep[PHI;][]{Solanki2020} on board SolO consists of two telescopes: the Full Disc Telescope (FDT) and the High Resolution Telescope (HRT).  
FDT observes the full solar disk across all orbital phases, capturing LoS magnetic field maps with a field of view of approximately 2$^{\circ} \times$ 2$^{\circ}$ and a pixel sampling of 3.75 arcseconds per pixel.

\subsection{Method to estimate the X-ray source height} \label{subsec:height_estimate}
The event under investigation occurred on 2022 August 28 within NOAA AR 13088 at the west solar limb (heliographic coordinates: S27$^{\circ}$, W87$^{\circ}$), as viewed from the Earth perspective by SDO. This event is also captured by SolO, which is 0.78 AU away from the Sun and had a longitudinal separation of 147.1$^{\circ}$ from the Earth (Figure \ref{fig:ovv}).
There were two M-class flares in close succession from the target AR (Figure \ref{fig:spec_fit_tl}a): an M6.7 flare followed by an M4.6 flare. 
According to GOES soft X-ray emission curves, the M6.7 flare peaked at 16:19 UT, followed by the M4.6 flare at 18:32 UT. 
This study focuses on the first flare only.
To investigate the height evolution of the SolO/STIX X-ray sources and their corresponding EUV structures in AIA, we developed a method to estimate the source heights above the solar surface.

\begin{figure*}[ht!]
	\centering\includegraphics[width=0.9\textwidth]{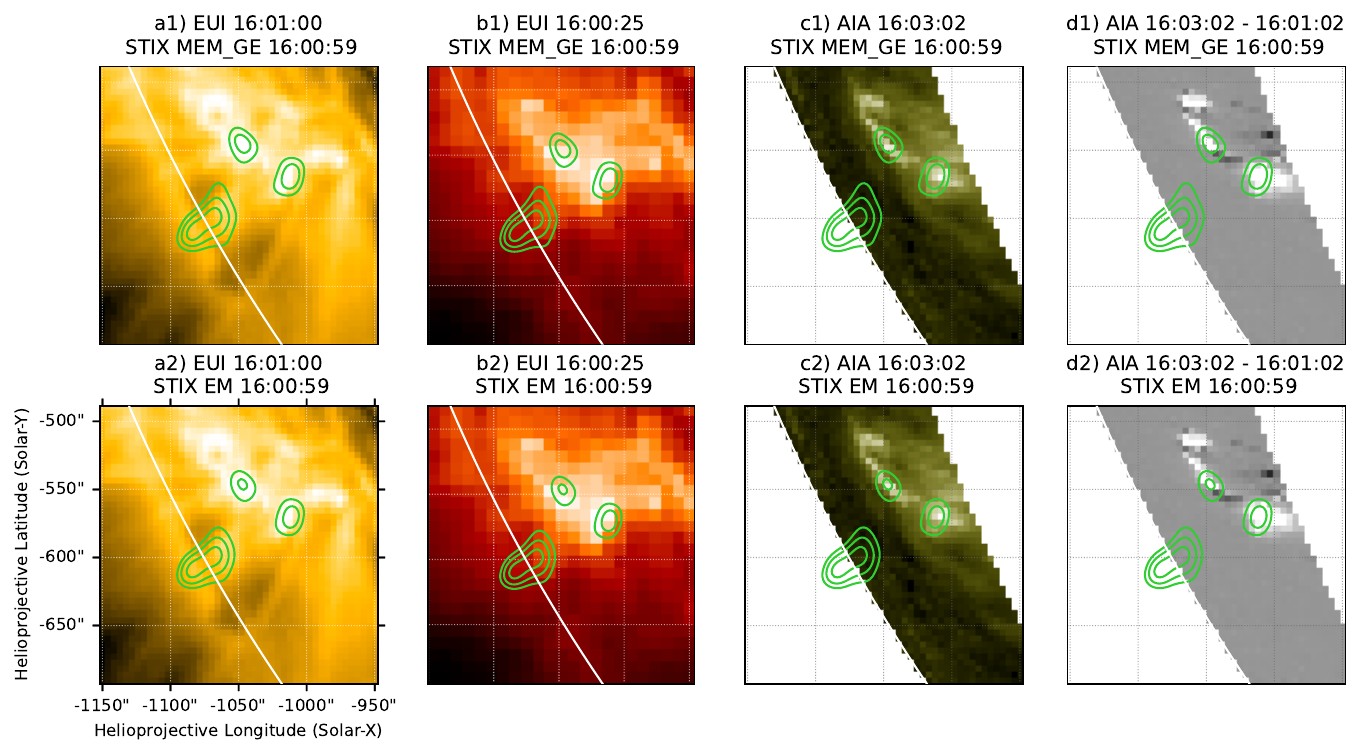}
	\caption{Co-alignment of the STIX X-ray source positions.
		The the background images from the left to right columns are EUI 174 and 304 \AA, reprojected AIA 1600 \AA~and corresponding difference image. The superimposed X-ray sources (16--28 keV) reconstructed by the MEM\_GE (1st row) and EM (2nd row) methods are shown as green contours (30\%, 50\%, 80\% of the maximum value). The time stamps are recorded by the instruments. 
		\label{fig:pos_calibration}}
\end{figure*}

Due to uncertainties in the instrumental pointing solution, reconstructed source positions from SolO/STIX data may contain offsets when the spacecraft is more than 0.75 AU from the Sun \citep{Massa2022}.
To manually correct this, we incorporated additional observational information during the preprocessing stage. 
Specifically, we reprojected an SDO/AIA 1600 \AA~image and its corresponding difference image into SolO's viewpoint (Figure \ref{fig:pos_calibration}c1–d1,c2-d2) using the procedure \texttt{reproject\_to} included in the SunPy package, which reprojects a map to a different world coordinate system (WCS).
Such a reprojection procedure requires only two input maps, their WCS information, the 2D position of the point in the original map and its distance from the solar center (derived from the estimated height), which can be very accurate, with the only sources of uncertainty arising from the resolution of the two maps and the positional error of the projected point.
Then we overplotted the footpoint sources onto four maps, i.e., SolO/EUI 174 and 304~{\AA}, SDO/AIA 1600~{\AA} and its corresponding difference image, taking into account the light traveling time difference of 115.24 sec between the Earth and SolO.
The STIX footpoint sources used in this comparison were reconstructed using two imaging reconstruction algorithms---MEM\_GE (Figure \ref{fig:pos_calibration}a1-d1) and EM (Figure \ref{fig:pos_calibration}a2-d2). 
The moment selected for reconstruction corresponds to the peak of the strongest non-thermal emission (16:03 UT in SDO/AIA time), when the clearest footpoint sources were visible.
We iteratively adjusted the STIX image to determine the optimal shift that best aligns the footpoint sources with the SDO/AIA 1600~{\AA} flare ribbons, while also considering the structures observed in SolO/EUI 174 and 304~{\AA} images. 
After weighing the overall co-alignment across all reference images, we determined the best-fit correction to be a shift of the STIX images by $-12$ arcseconds in the x-direction and $+40$ arcseconds in the y-direction, which is well within the $\pm$100 arcseconds pointing uncertainty reported by \citep{Massa2022}. 
This correction was applied to all STIX imaging results used in the subsequent analysis, as \citet{Massa2022} noted that the shift is expected to remain relatively consistent for individual flares.

To further confirm the plausibility of the co-alignment method, we overplotted the footpoint source onto the LoS magnetic field data from two instruments, SolO/PHI (Figure \ref{fig:fp_plus_B}a) and SDO/HMI (Figure \ref{fig:fp_plus_B}b-c).
Despite significant projection effects due to their vastly different perspectives (Figure \ref{fig:ovv}), both instruments revealed that the footpoint sources correspond to opposite magnetic polarities, i.e., on either side of the polarity inversion line (PIL) in SolO/PHI (marked as `PIL-A') and in SDO/HMI (marked as `PIL-B'), consistent with the standard flare model.
A polarity discrepancy arises in the region between the footpoints (orange arrows): a patch to the west of main positive polarity appears negative in HMI but remains positive in PHI.
This phenomenon can still be explained by considering the large difference in viewing angle between the two satellites (147.1$^{\circ}$): the magnetic fields connecting the two polarities are presumably low-lying and nearly horizontal, therefore projecting outward (toward the observer) in PHI's perspective, but inward (away from the observer) in HMI's.

\begin{figure*}[ht!]
	\centering\includegraphics[width=0.9\textwidth]{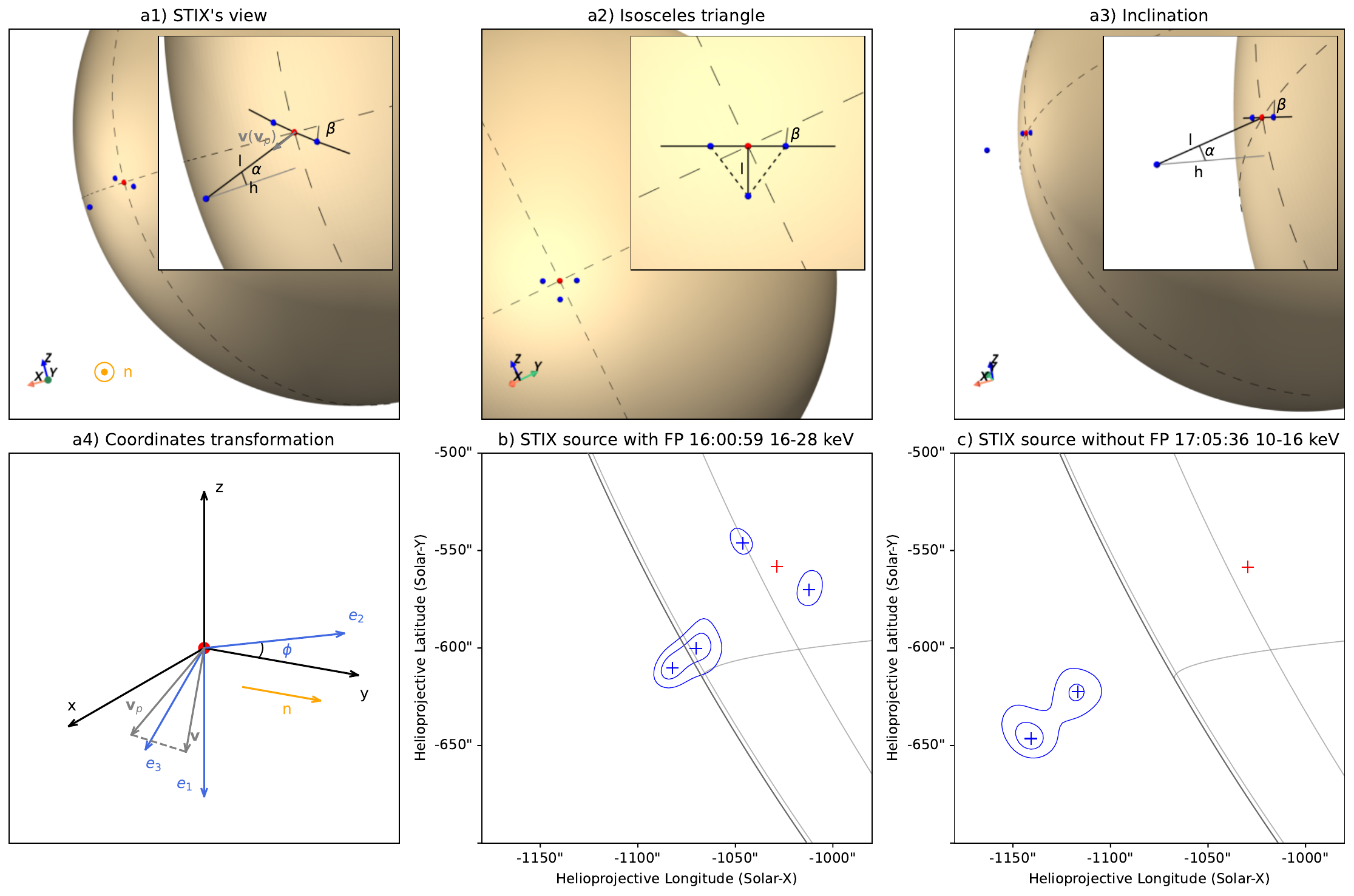}
	\caption{Method to estimate the height of coronal X-ray sources above the solar surface.
		Panels (a1–a4) illustrate the basic assumptions and coordinate systems to help determine the height of the loop-top source.
		The blue dots indicate the the observed X-ray sources (corresponding to the blue plus signs in panel (b)), while the red dot indicates the midpoint between the two footpoint sources (corresponding to the red plus signs in panel b). The unit vector $\mathbf{\hat{v}}$ points from the red to the blue dot above the surface, and its projection onto the observational plane is denoted by the unit vector $\mathbf{\hat{v}}_p$.	
		The $xyz$ axes in the coordinate system `O' are indicated by the arrows in the bottom left corner of panels (a1–a3), and the unit vector $\mathbf{\hat{n}}$ is perpendicular to the observation plane. The relationship between the coordinate systems `O' and `L' (see the text) is illustrated in (a4). Panels (b-c) provide examples of the observed coronal X-ray sources, with (panel (b)) or without (panel (c)) footpoint sources.
		\label{fig:ht_method}}
\end{figure*}

The reconstructed STIX 16--28 keV images reveals two footpoints and a double loop-top source. We identified the intensity maxima (marked by blue plus symbols in Figure \ref{fig:ht_method}b) and defined the midpoint (red plus) of the line connecting the two footpoint sources. These are denoted by the same colors in the schematic plots in Figure~\ref{fig:ht_method}(a1--a4).
In the following, based on four key positions---the two footpoints and their midpoint and one specified coronal source---we calculate the coronal source's height by applying the assumptions and definitions below:
\begin{itemize}
    \item \textbf{Assumption 1}: The two footpoint sources are fixed at the surface of the sun, i.e., at a height of $0$~Mm, while the loop-top source evolves above the surface.
    \item \textbf{Assumption 2}: An isosceles triangle is formed by the three sources, with equal distance between the coronal source and the two footpoints (Figure \ref{fig:ht_method} a2).
    \item \textbf{Definition 1}: $\alpha$ is defined as the inclination angle of the line connecting the coronal source (blue) to the footpoint midpoint (red) with respect to the direction vertical to the solar surface (Figure \ref{fig:ht_method}(a3)).
    \item \textbf{Definition 2}: $\beta$ is defined as the angle between the line connecting the two footpoints and the great arc connecting the red midpoint to the center of the solar disk (a gray dashed line in Figure \ref{fig:ht_method}a1). This angle can be directly measured from the positions of the two footpoint sources (Figure \ref{fig:ht_method}(a1-a3)).
    \item \textbf{Definition 3}: $l$ is the distance between the midpoint and the coronal source.
    \item \textbf{Definition 4}: $h$ is the vertical distance of the source with respect to the surface.
\end{itemize}

Given the location of the midpoint, we established two rectangular (Cartesian) coordinate systems and one spherical coordinate system (see panels a1 and a4 in Figure \ref{fig:ht_method}). The Cartesian coordinate systems are used to describe the relative positions of the sources, while the spherical system represents their positions on the solar surface.
Both Cartesian coordinate systems are centered at the red midpoint.

The first Cartesian coordinate system, referred to as the `coordinate system O', meaning `observational', is defined as follows : The origin is the midpoint of the line connecting two footpoint sources (red point in Figure \ref{fig:ht_method}a1-a3);
The $y$-axis (green in Figure \ref{fig:ht_method}a1–a3) points along the line of sight toward the observer;
The $x$-axis lies in the observational (image) plane, pointing from the center of the Sun toward the red midpoint.
The $z$-axis completes the right-handed system, also lying in the observational plane and perpendicular to the x-axis.
The $xyz$ directions are illustrated by the orange, green, and blue arrows, respectively, in Figure \ref{fig:ht_method}(a1--a3).

The origin of the spherical coordinate system is the center of the Sun. The zenith angle $\theta$ is defined as the angle between the position vector of a point on the spherical surface and the $z$-axis.  
The azimuth angle $\phi$ is the angle between the projection of this position vector onto the $xy$-plane and the $x$-axis.  
With this definition, the red midpoint corresponds to $\theta = 0^{\circ}$, and its azimuth angle $\phi$ can be computed from its observed position on the solar surface using the formula: $\cos \phi = \frac{d}{R_{\astrosun}}$, where $R_{\astrosun}$ is the solar radius and $d$ is the projected distance from the midpoint to the disk center in the observational plane.

The second Cartesian coordinate system, referred to as the `coordinate system L', meaning `local', is defined using the unit vectors $(\mathbf{\hat{e}}_1, \mathbf{\hat{e}}_2, \mathbf{\hat{e}}_3)$ as its base axes and has the same origin as the coordinate system O. Its relationship to the coordinate system O, with base vectors $(\mathbf{\hat{e}}_x, \mathbf{\hat{e}}_y, \mathbf{\hat{e}}_z)$, is shown in Figure \ref{fig:ht_method}(a4) and expressed by the following matrix:
\begin{equation}
	\begin{pmatrix}
		\mathbf{\hat{e}}_x \\
		\mathbf{\hat{e}}_y \\
		\mathbf{\hat{e}}_z
	\end{pmatrix}
	=
	\begin{pmatrix}
		0 & -\sin\phi & \cos\phi \\
		0 & \cos\phi & \sin\phi \\
		-1 & 0 & 0
	\end{pmatrix}
	\begin{pmatrix}
		\mathbf{\hat{e}}_1 \\
		\mathbf{\hat{e}}_2 \\
		\mathbf{\hat{e}}_3
	\end{pmatrix},
\end{equation}
where $\mathbf{\hat{e}}_3$ is directed along the local radial direction at the red midpoint. $\mathbf{\hat{e}}_1$ and $\mathbf{\hat{e}}_2$ respectively follow the local directions of $\mathbf{\hat{e}}_{\theta}$ and $\mathbf{\hat{e}}_{\phi}$ in the spherical coordinate system.

In the coordinate system L, the unit vector $\mathbf{\hat{v}}$ pointing from the footpoint midpoint to the coronal source is expressed as
\begin{equation}
	\mathbf{\hat{v}}^{L}=(\hat{v}_{1}, \hat{v}_{2}, \hat{v}_{3}) =(\sin \alpha \cos \beta, -\sin \alpha \sin \beta, \cos \alpha).
\end{equation}
To transform $\mathbf{\hat{v}}$ to the coordinate system O, we use the following transformation:
\begin{equation}
	\mathbf{\hat{v}}^{O} =
	\begin{pmatrix}
		0 & -\sin\phi & \cos\phi \\
		0 & \cos\phi & \sin\phi \\
		-1 & 0 & 0
	\end{pmatrix}
	\begin{pmatrix}
		\hat{v}_{1} \\
		\hat{v}_{2} \\
		\hat{v}_{3}
	\end{pmatrix}.
	\label{eq:vtop_cartesian}
\end{equation}
The unit vector perpendicular to the observational plane (the orange vector in Figure \ref{fig:ht_method} (a1 \& a4)) is $\mathbf{\hat{n}}=(0,1,0)$ in the coordinate system O. Therefore, the projection of $\mathbf{\hat{v}}$ onto the observational plane is given in the coordinate system O as (below the superscript O is dropped for simplicity):
\begin{align}
	\mathbf{\hat{v}}_{p} 
	&= \mathbf{\hat{v}}- (\mathbf{\hat{v}} \cdot \mathbf{\hat{n}}) \mathbf{\hat{n}} = 
	\begin{pmatrix}
		\cos \alpha \cos \phi + \sin \alpha \sin \beta \sin \phi \\
		0 \\
		-\sin \alpha \cos \beta
	\end{pmatrix}.
	\label{eq:vplane}
\end{align}
Projecting the vector $\mathbf{v}$ connecting the midpoint and the coronal source onto the observational plane gives
\begin{align} \label{eq:lvplane}
\mathbf{v}_{p} & = (l_x,\, 0,\, -l_z) \nonumber\\
               & = (l(\cos\alpha \cos \phi + \sin\alpha\sin \beta\sin\phi),\, 0,\, -l\sin\alpha \cos \beta),
\end{align}
in which $l_{x}$, $l_{z}$, $\beta$, and $\phi$ can be obtained by observation. Hence, with Eq.~(\ref{eq:lvplane}), we take the ratio $l_x/l_z$ to solve
for $\cot\alpha$:
\begin{equation}
	\cot \alpha = \frac{l_x}{l_z}\frac{\cos\beta}{\cos\phi}-\sin\beta\tan\phi;
\end{equation}
and then $h$ is given as follows:
\begin{equation}
	h=l\cos \alpha=l_z\frac{\cot \alpha}{\cos \beta}\\
	=\frac{l_x}{\cos \phi}-l_z\tan\beta\tan\phi.
\end{equation}

Since no discernible footpoint sources were observed for certain time intervals, we opted to use a fixed position as the footpoints' ``midpoint'' for all images. Specifically, we reconstructed images at 16:01 UT (SolO time) using two algorithms (MEM\_GE and EM), performed initial corrections, and identified the midpoint in each image. The average of these two positions was then calculated and adopted as the fixed midpoint.
We also considered the effect of solar rotation on the true position of the midpoint. 
However, this effect was found to be negligible within a three-hour window near the solar limb, so no adjustment was applied.
This fixed midpoint is applied to estimate the heights of all coronal sources in this event.

\section{Results} \label{sec:obs}
\subsection{Event overview}
\begin{figure*}[ht!]
	\centering\includegraphics[width=0.9\textwidth]{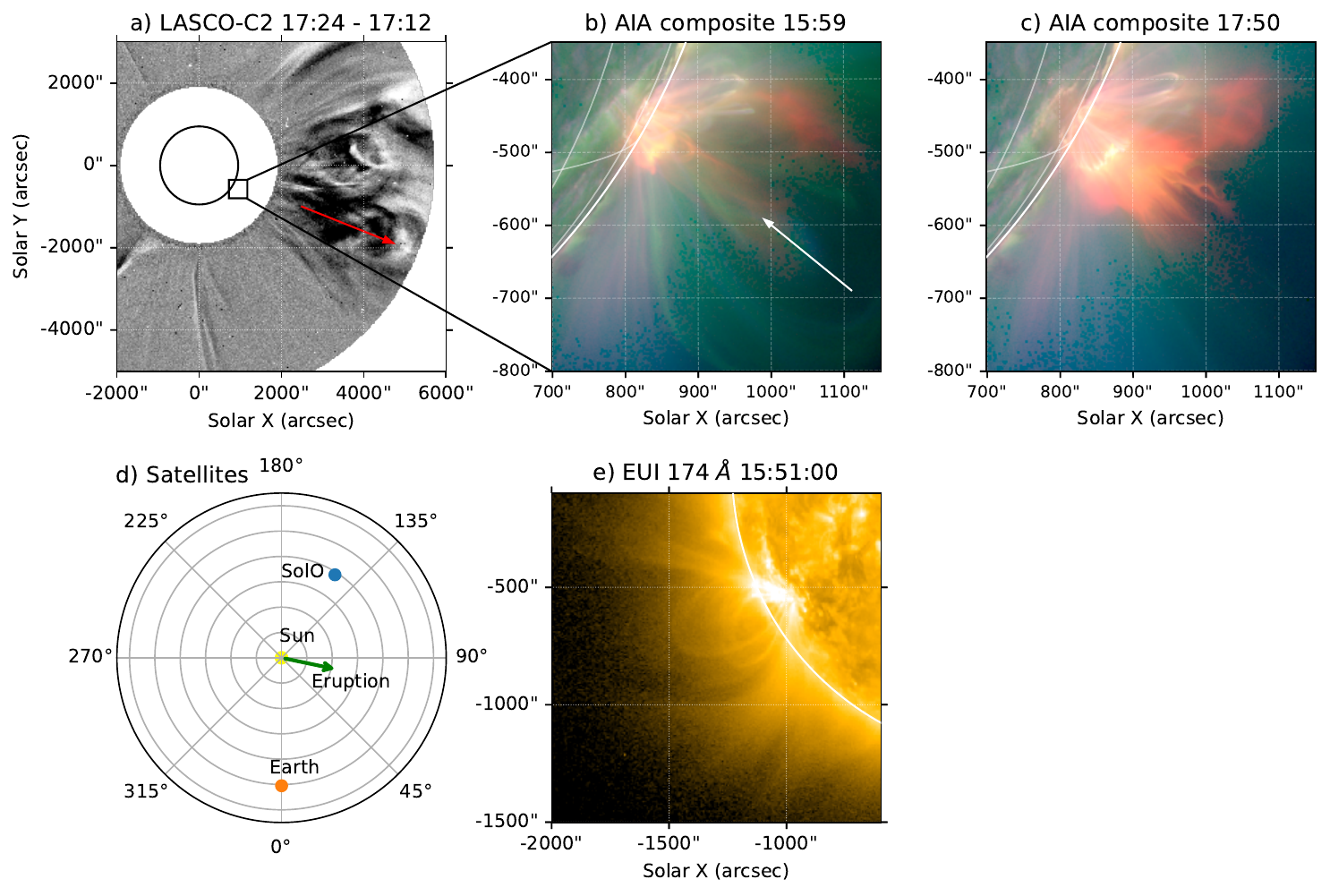}
	\caption{Overview of the eruption from two different viewpoints of the SDO and SolO spacecrafts. The red arrow in panel (a) indicates the CME detected by SOHO/LASCO C2. Panels (b-c) show the snapshots before and after the associated M7.6 flare by combining the three AIA passbands, 131 (red), 193 (green), and 171~{\AA} (blue), with the white arrow indicating the twisted, tube-like structure. The field of view in panels (b-c) corresponds to the small black rectangle in (a). In panel (d) the blue, orange and yellow dots indicate the relative positions of the satellites SolO, SDO (Earth) and the Sun, respectively, and the green arrow indicates the direction of the eruption. Panel (e) shows the active region in the EUI 174 \AA~image near the start time of the flare. The time stamps shown in each panel correspond to the original observation times recorded by the respective instruments. The accompanying animation shows the X-ray light curves in the same format as in Figure~\ref{fig:spec_fit_tl}a, the time-elapsed AIA 131, 193, and 171~{\AA} images, as well as the composite images combining the three narrow passbands, from 15:30 to 19:15 UT on 2022 August 28. 	\label{fig:ovv}}
\end{figure*}

During the impulsive phase of the M6.7 flare, a complex eruption was observed, involving a twisted, tube-like structure in the hot 131~{\AA} passband (red colors in Figure \ref{fig:ovv}b) and a group of overlying coronal loops visible in cooler passbands such as 171 {\AA} (blue) and 193 {\AA} (green; see also the accompanying movie).  
During the decay phase, hot, diffuse SAF became visible above the post-flare arcade (Figure \ref{fig:ovv}c). 
In the late decay phase of the flare, a series of SADs appeared, overlapping partially in time with the subsequent M4.6 flare. 
At 16:12 UT, approximately 20 minutes after the onset of the M6.7 flare, a coronal mass ejection (CME) was detected by the SOHO/LASCO coronagraph \citep{Brueckner1995}, confirming the eruptive nature of the event. 
Figure \ref{fig:ovv}(a) shows a LASCO C2 base difference image that highlights the CME (indicated by a red arrow), to the north of which another CME-like structure was visible. 
However, this second CME results from a slowly evolving coronal streamer that had appeared about five hours earlier. 
According to the event catalog provided by the SOHO/LASCO team, the second CME was back-extrapolated to a launch time several hours prior to the M6.7 flare. This indicates that it was unrelated to the eruption under investigation and likely originated from a separate, earlier event.

The M6.7-class flare exhibited prolonged X-ray emission, characterized by a main phase lasting approximately 30 minutes and a decay phase extending over more than two hours, temporally overlapped with the subsequent M4.6-class flare. 
The X-ray spectra of the M6.7-class flare evolved significantly over time (Figure \ref{fig:spec_fit_tl}a). According to the spectral fitting results, clear non-thermal components were shown in the first two time intervals (Figures \ref{fig:spec_fit_tl}b-c), with thick-target electron spectral indices of $\delta=$ 7.4 and 6.0, respectively.  
The spectrum hardened from the first interval to the second one, which is consistent with the fact that the footpoint sources (contours in Figure \ref{fig:pos_calibration}), typically associated with strong non-thermal emission, appeared only in the second interval.  
Around 16:01 UT (SolO time), in addition to the thermal component at a temperature of about 16 MK, a second thermal component at a super hot temperature of about 30 MK might be present. After the impulsive phase ended, the X-ray emission is again dominated by thermal emission (Figures \ref{fig:spec_fit_tl}d-f), as the spectral indices would be as large as over 10 if a thick-target model were used for fitting.

\subsection{X-ray source evolution}
At the onset of the M6.7 flare, X-ray imaging revealed at least three distinct sources (Figure \ref{fig:sc_evo}b1-b4): a dominant loop-top source and two weaker, transient footpoint sources. 
The footpoint sources were detectable exclusively in the 16--28 keV energy range during the impulsive phase of the flare (15:50--16:10 UT, SolO time), while the loop-top source persisted longer across both thermal and non-thermal emissions (Figure \ref{fig:sc_evo}). 
The loop-top source evolved differently in different energy ranges.
In the 16--28 keV range, the loop-top source as a whole remained discernible until 16:45 UT. 
It briefly fragmented into a double-source structure near the non-thermal emission peak (16:01 UT in SolO time, Figure \ref{fig:sc_evo}b1-b4).
In lower energy ranges (4--16 keV), a double source emerged minutes after the flare onset and lasted for an extended duration of about 20--30 minutes (Figure \ref{fig:sc_evo}c-d). 
Additional faint sources sporadically appeared in the 4--6 keV and 6--10 keV ranges later, though these were significantly weaker than the primary double source. 
The sources finally merged into one coronal source again, located tens of Mm above the solar surface, and persisted throughout the decay phase until the onset of the subsequent M4.6 flare (Figure \ref{fig:sc_evo}c-e).

\subsubsection{Thermal properties of the X-ray coronal emission}\label{subsubsec:thermal}
By applying the height estimation method to each pixel along the X-ray source contour (defined by the 50\% maximum intensity), we projected the source into the AIA field of view using the procedure \texttt{reproject\_to} included in the SunPy package, and compared the source with corresponding EUV structures. Based on the DEM analysis of AIA EUV images, we can also derive the properties of the hot EUV emitting coronal plasma and its spatio-temporal relation to the coronal STIX HXR sources (Figure \ref{fig:sc_evo}). 

Ten minutes before the eruption, the region containing the X-ray source was filled with a dense set of coronal loops, exhibiting temperatures around $6 \times 10^6$ K and an emission measure of approximately $10^{28}$~cm$^{-5}$ (Figure \ref{fig:sc_evo}(a1–a5)).  
At the peak of STIX non-thermal emission (16--28 keV), a double coronal source (red contours) became prominent at the loop top, coinciding with the eruption of the magnetic flux rope (Figure \ref{fig:sc_evo}(b1–b5)). By averaging DEM-weighted temperature, EM and $T$-EM distribution within two black rectangular regions (Figure \ref{fig:sc_evo}(c3--c5, d3--d5)) which are placed approximately at the central regions of different sources (also taking into account the relative offsets of the source centroids observed in different energy ranges), we found that the upper source is situated in a region of higher temperature ($\sim$20.6 MK) and higher emission measure ($\sim 4.9 \times 10^{29}$ cm$^{-5}$), while the lower source in a region of lower temperature ($\sim$13.8 MK) and lower emission measure ($\sim 3.5 \times 10^{29}$ cm$^{-5}$). This lower source is also co-spatial with the sources at lower energies (4--16 keV). Accordingly, from the $T$-EM distribution (5th row in Figure \ref{fig:sc_evo}), one can see that the hot component of the upper source (orange) peaks at $\log_{10}T\approx7.2$, while the counterpart of the lower source (blue) peaks at $\log_{10}T\approx7.0$. The presence of multi-thermal, X-ray-emitting plasmas at the loop top is consistent with the X-ray spectrum observed at the same time (Figure~\ref{fig:spec_fit_tl}c), as the spectrum is best fit by two thermal components of temperatures at about 16 and 31 MK, respectively, plus the nonthermal component. The temperature differences between the DEM and spectral fitting results may be attributed to different temperature response functions of AIA and STIX, with the former being less sensitive to $>20$ MK plasmas. The nonthermal component is attributed to the two footpoint sources, though less intense than the coronal source, were visible and roughly projected onto the ends of the coronal loops. 

As time progressed, the STIX X-ray sources became more diffuse and spread over a larger area, particularly in the 4--10 keV range (Figure \ref{fig:sc_evo}(c1–c5)). The 4--10 keV source exhibited a broad base associated with low temperatures but high emission measure, and a narrower top associated with high temperatures and low emission measure. During this period, several spike-like thermal structures emerged in the loop-top region, most prominently in the AIA 131 \AA~images and temperature maps (Figure \ref{fig:sc_evo}(c2 \& c3)).  
Occasionally, a 16--28 keV source appeared above the most pronounced spike (marked by an arrow in Figure \ref{fig:sc_evo}(c2 \& c3)), suggesting that the spikes might be associated with the energy release above the loop top \cite[]{Liu&Wang2021}. 
Approximately an hour after the flare onset, a double source structure reappeared in the 4--16 keV range, resembling the earlier 16--28 keV configuration at 16:03 UT: the higher source was hotter with lower emission measure, and the lower source was cooler with higher emission measure (Figure \ref{fig:sc_evo}(d1-d5)).  
Eventually, the two sources merged into a single, larger structure (Figure \ref{fig:sc_evo}(e1-e5)).

\subsubsection{Multi-phase evolution of the X-ray coronal emission} \label{subsubsec:ht_evo}
We applied the height estimation method (\S\ref{subsec:height_estimate}) to a set of well-detected coronal sources with rich evolutionary features.
Figure~\ref{fig:ht_evo_mem_ge} shows the obtained height-time evolution of the STIX sources derived in four different energy bands, along with STIX light curves.
The observational times of all STIX sources are corrected for the light traveling time from SolO to Earth (115.24 seconds). Source positions are determined by the local maxima within the 50\% contour of the reconstructed X-ray images. 
The figures show the time evolution of the estimated heights of these identified sources.
To filter out sources that are too faint, we define a threshold of intensity value, $f_{\text{th}}$, for each energy range.  
This threshold is given by $f_\mathrm{th} = f_{\min} \times 10^{(\log_{10}f_{\max} - \log_{10}f_{\min})\,k_\mathrm{th} }$, where $k_\mathrm{th}$ is a tuning parameter, and $f_{\min}$ ($f_{\max}$) refers to the minimum (maximum) intensity in the reconstructed X-ray image. In this study, $k_\mathrm{th}$ is set to be 0.2 through a trial-and-error approach to single out robust sources.  
Data points with peak intensities below the threshold are shown in gray in Figure~\ref{fig:ht_evo_mem_ge}, while others are represented by circles whose radii indicate the maximum source intensity. This filtering procedure is applied throughout the subsequent analysis, including the linear regression of the source height-time evolution.

The height-time plots reveal several clearly distinguishable clustered sources, each showing distinct evolutionary trends. We grouped the data points into `A', `B', `C', `I', or `O',  according to their height evolution in X-ray images reconstructed by three different methods, i.e., MEM\_GE, EM, and CLEAN. 
The following discussion is based primarily on MEM\_GE imaging reconstructions (Figure \ref{fig:ht_evo_mem_ge}), which give similar results as the EM and CLEAN methods (not shown).

In the 4--6, 6--10, and 10--16 keV energy ranges, at least three distinct groups, `A', `B', `C', are evident, corresponding to three different sources.  
Source A (pink dots in Figure \ref{fig:ht_evo_mem_ge}; also see Figure \ref{fig:sc_evo}a–b) shows an initial downward motion (7--15~km~s$^{-1}$) beginning at flare onset, followed by a gradual upward trend (3--7~km~s$^{-1}$).  
The early downward motion has been widely studied \citep{Sui2004,OFlannagain2013,Ji2007,Shen2008,Veronig2006,Unverferth2023,Li2006}, with explanations including spectrum hardening \citep{OFlannagain2013}, magnetic shear release \citep{Ji2007}, post-reconnection flux tube contraction \citep{Li2006,Shen2008,Unverferth2023}, and plasma heating in a collapsing magnetic trap \citep{Veronig2006}.  
Source B (brown dots in Figure \ref{fig:ht_evo_mem_ge}; also see Figure \ref{fig:sc_evo}b–d) emerges around 16:30 UT. Its rising speed in the 6--10 keV range ($\sim$10~km~s$^{-1}$) is notably faster than in the 4--6 and 10--16 keV ranges (3--6~km~s$^{-1}$).  
Source C (cyan dots in Figure \ref{fig:ht_evo_mem_ge}; also see Figure \ref{fig:sc_evo}b–d) appears at 4--6 keV around 16:30 UT and later at 6--10 and 10--16 keV around 16:55 UT, rising at 2--3~km~s$^{-1}$.

The 16--28 keV sources last for a shorter duration, until about 16:45 UT (Figure \ref{fig:ht_evo_mem_ge}e), but two evolutionary phases are still distinguishable: earlier sources are stronger and more compact (pink dots; labeled `A') than later ones which are weaker and more dispersed (brown dots; labeled `B'). Again, the initial downward motion is evident in all three algorithms: 24~km~s$^{-1}$ in MEM\_GE, 14~km~s$^{-1}$ in EM, and 10~km~s$^{-1}$ in CLEAN. At the non-thermal emission peak, a double loop-top structure is present (Figure \ref{fig:sc_evo}b), and collectively the source height decreases at a speed of $\sim$4~km~s$^{-1}$. The dispersed source `B' in the second phase rises at $\sim$11~km~s$^{-1}$ and fades at around 16:45 UT. 

\begin{figure*}[ht!]
	\centering\includegraphics[width=0.9\textwidth]{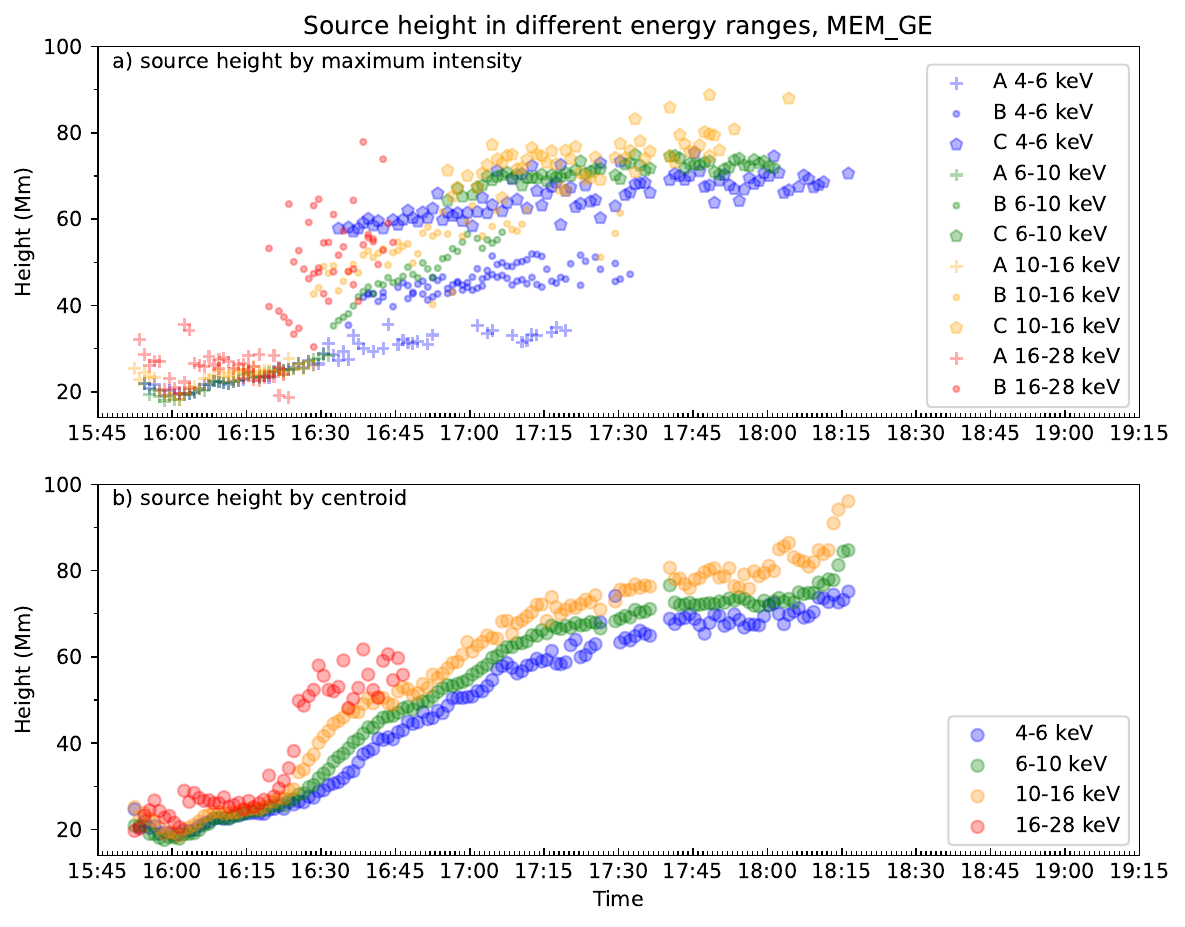}
	\caption{STIX X-ray source heights determined by maximum intensity versus centroid. The X-ray sources are reconstructed with the MEM\_GE method. Panel (a) shows the source heights determined by the location of maximum intensity within the 50\% contours. The four colors indicate the different energy ranges, and the three symbols (`+', circle, pentagon) correspond to the three groups of sources: A, B, and C. Panel (b) shows the source heights by their centroid (intensity-weighted) positions. The times have been corrected for the light traveling from SolO to Earth. } 	\label{fig:ht_distr_mem_ge}
\end{figure*}

For a certain period, the coronal X-ray sources B and C coexisted and dominated the X-ray emission. The exact start and end times of this coexistence varied with energy. In 4--6 keV, it lasted from 16:40 UT to 17:40 UT; in 6--10 keV, from 16:55 UT to 17:07 UT; and in 10--16 keV, from 16:55 UT to 17:12 UT (Figure \ref{fig:ht_distr_mem_ge}).

To further investigate the source height evolution as a function of energy, we plotted all the points in each energy range whose intensities exceed the threshold in Figure \ref{fig:ht_distr_mem_ge}, using different colors to differentiate energy ranges and symbols to differentiate sources, i.e., A, B, and C. The source height is either determined by the maximum intensity (Figure \ref{fig:ht_distr_mem_ge}a) or the intensity-weighted centroid position (Figure \ref{fig:ht_distr_mem_ge}b). A clear trend can be seen in the distribution of Sources B (circles) and C (pentagons): higher-energy sources tend to be located at higher altitudes. 
This energy-height distribution differs from previous results on double coronal sources \citep[e.g.,][]{Sui2003, Liu2008}, which typically show that the lower source exhibits higher-energy emissions at greater altitudes, whereas the upper source displays the reversed pattern.

\section{Discussion and Conclusion} \label{sec:conclusion}
In summary, we investigated the detailed evolution of the X-ray sources over a period of approximately two hours during an M6.7-class flare. At least three relatively independent sources were identified, each displaying distinct dynamic patterns and undergoing multiple evolutionary stages.
We found that the spatial distribution of the sources varies with energy: higher-energy sources are located at higher altitudes. 
This trend is generally consistent with the behavior of flare loop-top sources and can be attributed to the rising reconnection site in eruptive flares and the consequent cooling of flare loops, resulting in hotter flare loops formed at greater heights in the corona, with increasingly cooler loops nested underneath \citep[e.g.,][]{Veronig2006,Vrsnak2006}.

\begin{figure*}[ht!]
	\centering\includegraphics[width=0.8\textwidth]{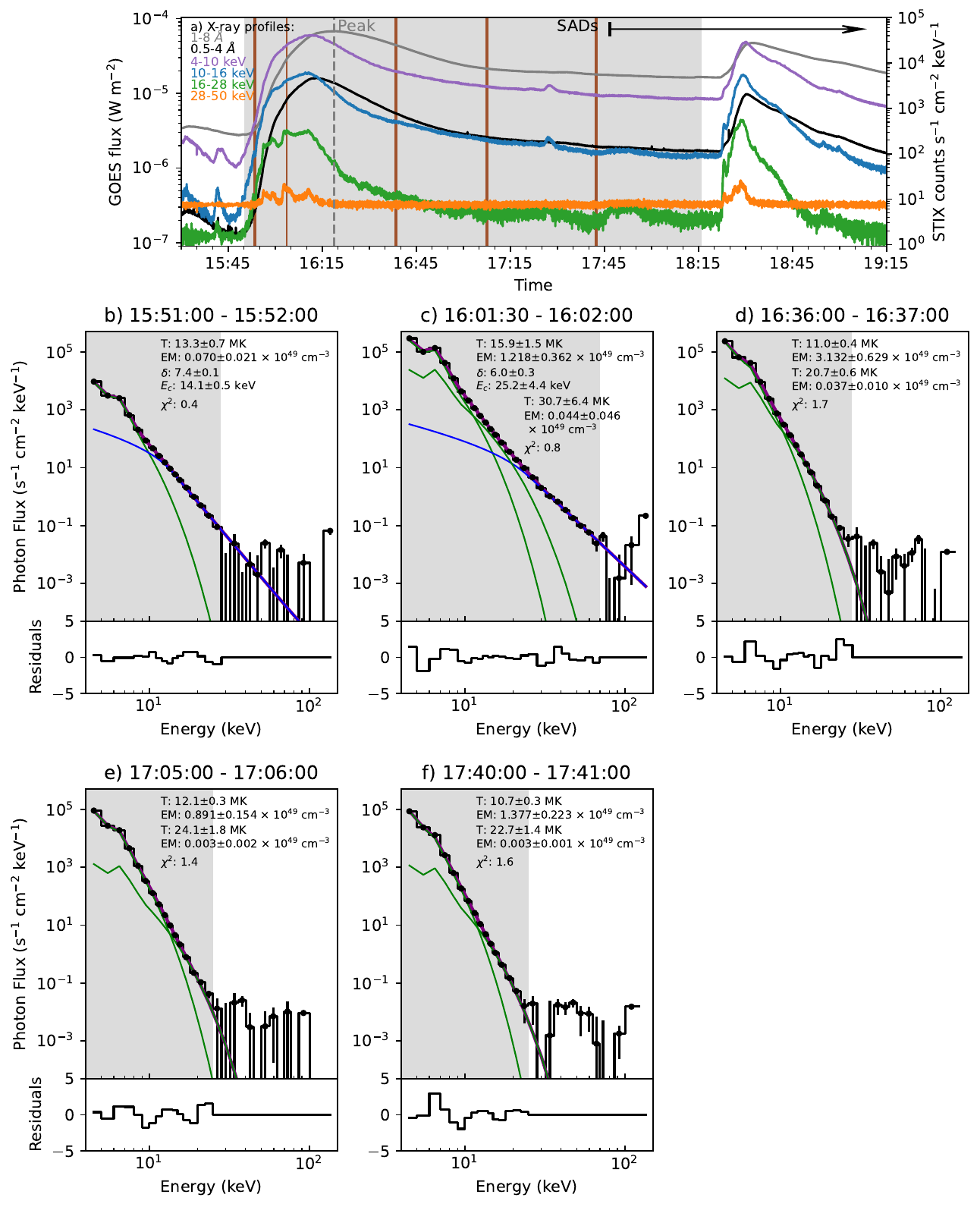}
	\caption{X-ray light curves and spectral fitting.
		In panel (a), the black and gray curves show respectively 1--8 \AA~and 0.5--4 \AA~soft X-ray flux observed by GOES. The colored curves show the X-ray count fluxes measured by SolO/STIX at different energy ranges from 4 to 50 keV. The original STIX observation times have been added by 115.24 seconds for the light traveling from the SolO to the Earth. The brown vertical lines indicate the corrected times of panels (b--f), in which the STIX time stamps are shown. The gray vertical line denotes the flare peak time. The horizontal long arrow marks the time interval during which SADs are observed. In panels (b--f), the observed photon fluxes are shown in histograms. The green and blue curves denote respectively the thermal and nonthermal component of the fitting scheme; the latter adopts the thick-target model. The sum of all the individual components is indicated by the purple curve.
		\label{fig:spec_fit_tl}}
\end{figure*}

\begin{figure*}[ht!]
	\centering\includegraphics[width=0.9\textwidth]{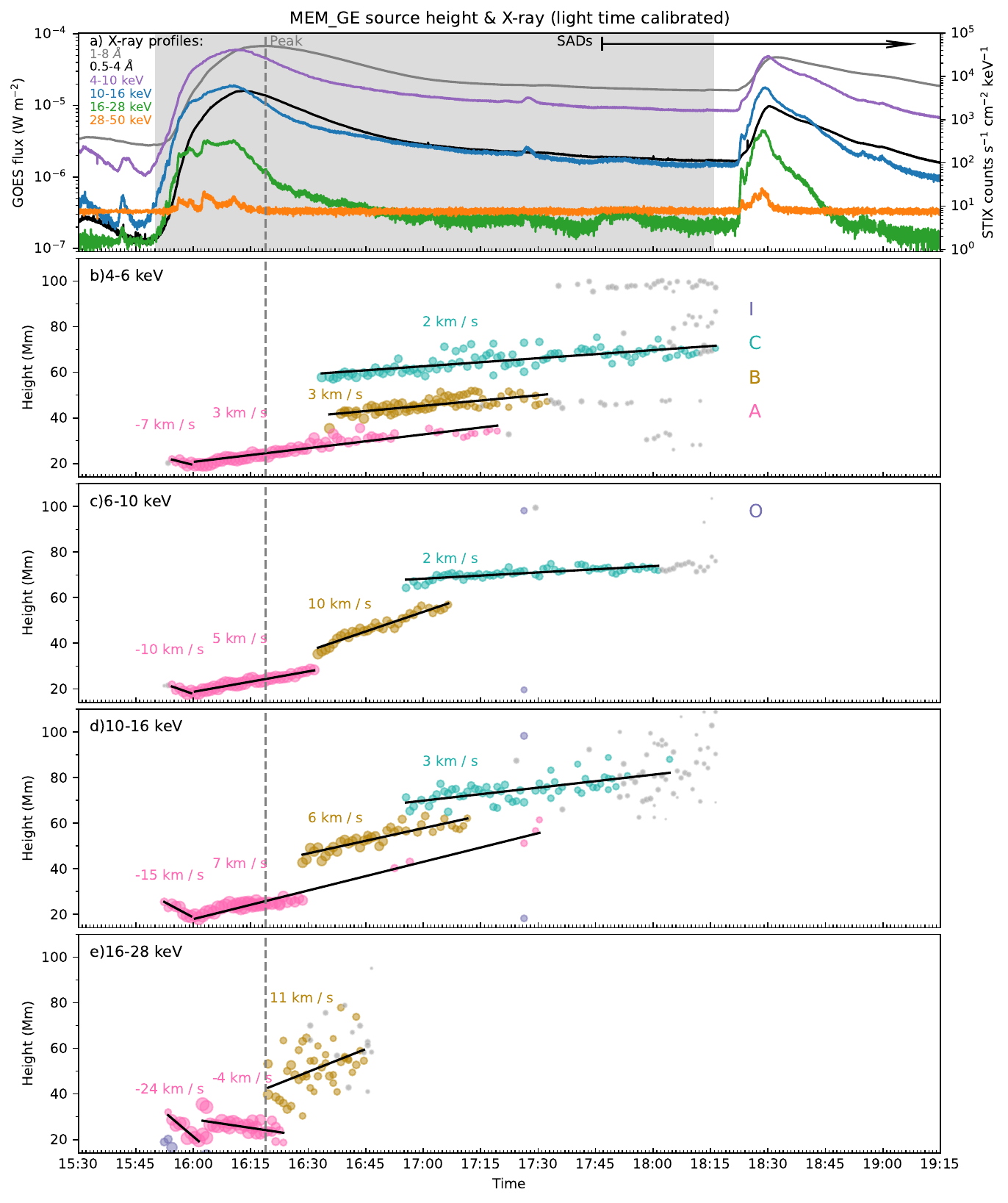}
	\caption{Height-time plots of STIX X-ray sources derived with MEM\_GE in four different energy bands. Panel (a) is identical to Figure \ref{fig:spec_fit_tl}(a). Panels (b-e): The colored dots represent the estimated heights of STIX X-ray sources detected at the corresponding times and energy ranges. The sizes of the dots reflect the relative intensities of the sources. Different groups of dots are marked by different colors and the letters `A', `B', `C', `I' and `O'. Gray dots indicate sources with intensities below a certain threshold and are excluded from the linear fitting (black lines) used to calculate the mean speeds of the sources. The times have been corrected for the light traveling from SolO to Earth. } \label{fig:ht_evo_mem_ge}
\end{figure*}

What is particularly noteworthy in this event is that such a energy-height correlation is also observed in each of the double X-ray coronal source, In contrast, previous studies \citep{Sui2003,Liu2008} typically show that the energy-height distribution of the two coronal sources mirror each other, with higher energy emission located closer to the center of the double source, where the energy is supposedly released.
In our case, the DEM results indicate that the double source corresponds to plasmas with different temperatures and densities (Section \ref{subsubsec:thermal}), consistent with the X-ray spectral fitting which requires two distinctive thermal components in addition to the nonthermal component (Figure~\ref{fig:spec_fit_tl}c). 
Due to the breaking of symmetry, the two coronal sources are unlikely located at the opposite sides of the reconnection point as suggested by \cite{Liu2008}, but both beneath the energy release site, since during the decay phase the reconnection point must have risen to a very high altitude. The fact that they co-exist for a certain period but are associated with plasmas with different temperatures might even suggest the presence of two spatio-temporally different energy release sites.  

\begin{figure*}[ht!]
	\centering\includegraphics[width=0.9\textwidth]{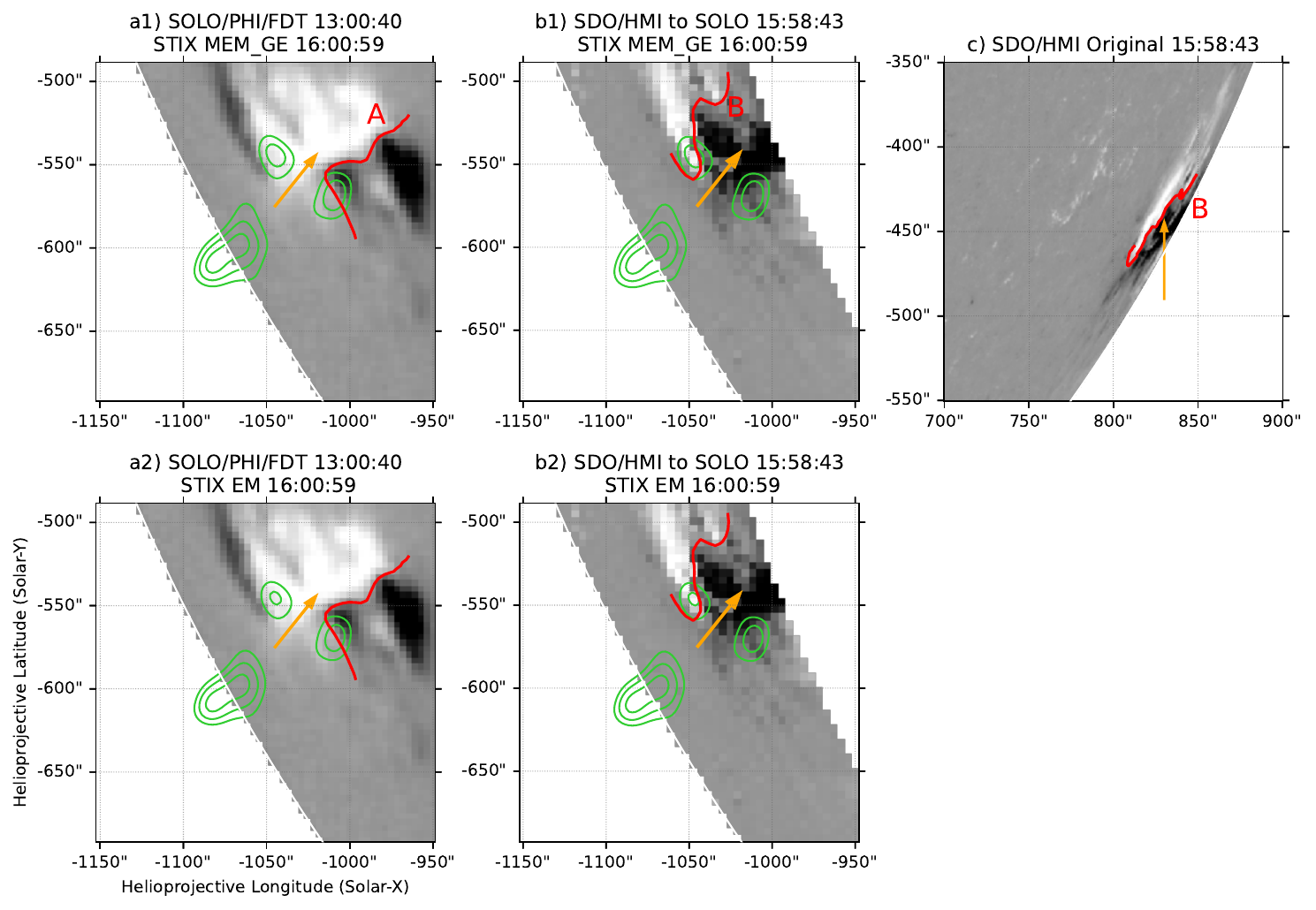}
	\caption{Photospheric magnetic field in relation to the STIX X-ray sources.
		Panles (a1--a2) show the line-of-sight (LoS) magnetic field observed by SolO/PHI. In panels (b1--b2) the LoS magnetic field observed by SDO/HMI is reprojected to the SolO perspective. The original HMI LoS magnetogram is shown in panel (c). Curves A and B denote the main PILs. 
		The orange arrow indicates where the magnetic field polarity is uncertain.
		The STIX X-ray sources (16--28 keV) obtained using the MEM\_GE and EM methods are superimposed as contours (30\%, 50\%, and 80\% of the maximum value). The time stamp shown in each panel is recorded by the respective instrument.
		\label{fig:fp_plus_B}}
\end{figure*}

\begin{figure*}[ht!]
	\centering\includegraphics[width=1.0\textwidth]{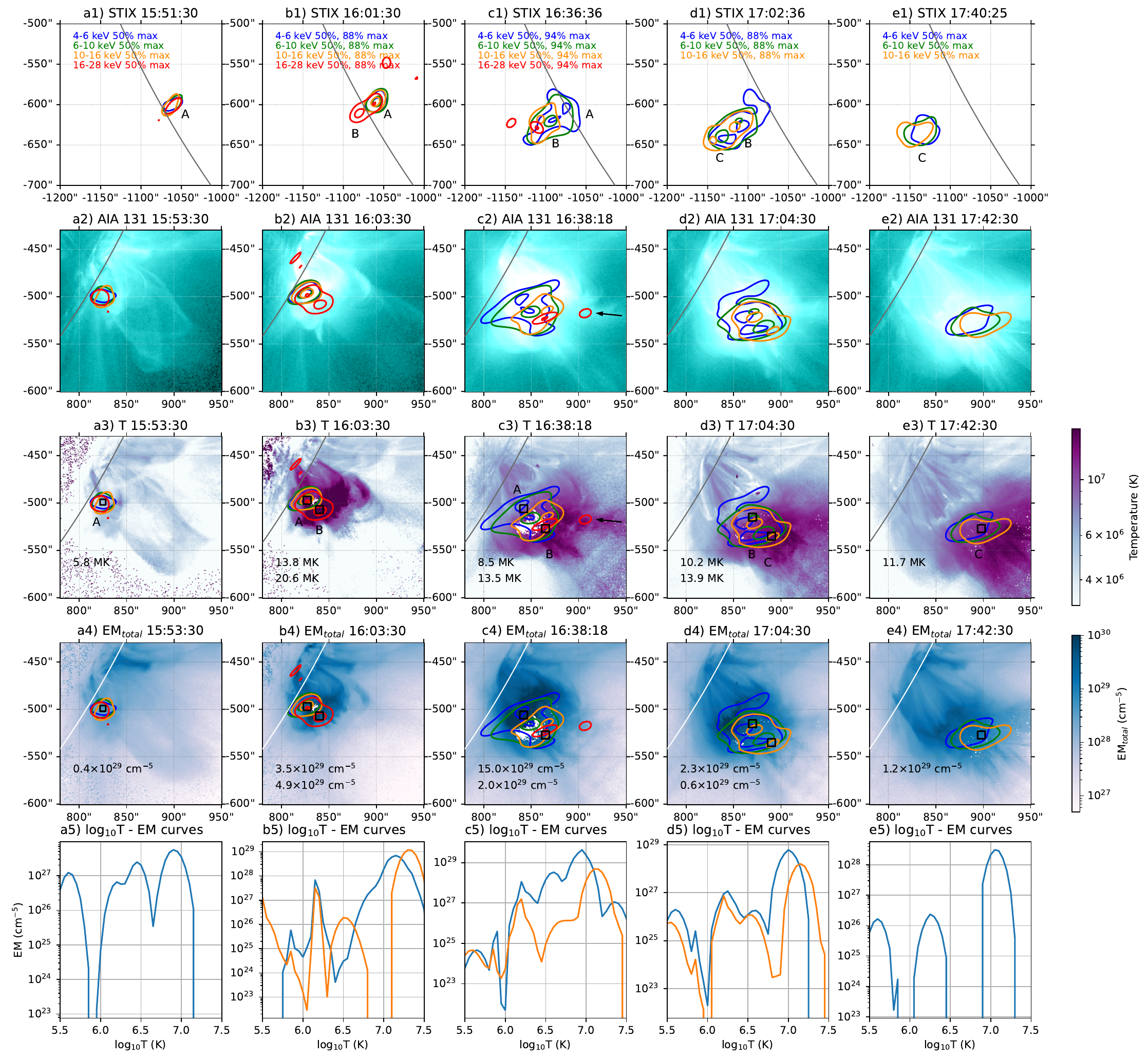}
	\caption{Evolution of X-ray sources. The STIX X-ray sources reconstructed by the MEM\_GE method at different energy bands from 4 to 28 keV (1st row) are reprojected to the perspective of SDO (2nd--4th rows). The re-projection was made with the estimated source heights (Section \ref{subsec:height_estimate}, Figure \ref{fig:ht_method}) and by the Python method \texttt{reproject\_to()} included in the SunPy package. Note that the contour levels are set as 50\%, 88\% or 94\% of maximum value for a clearer display of source structures, different from those in Figures \ref{fig:fp_plus_B} and \ref{fig:ht_method}. The X-ray sources are superimposed upon AIA 131~{\AA} images (2nd row), maps of DEM-weighted average temperature (3rd row), and maps of emission measure (4th row). The black arrow in panel (c2-c3) indicates the hot spike where a weak 16--28 keV source is located. The time stamp shown in each panel corresponds to the original observation time recorded by the respective instrument. 
    The black boxes in the 3rd--4th rows are placed approximately at the central regions of different sources; their locations are chosen by taking into account the relative offsets of the source centroids observed in different energy ranges. The EM-averaged temperature and the mean EM in these boxes are shown in the 3rd--4th rows, respectively. The $T$-EM distributions derived from the box regions are plotted in the 5th row, with EM [cm$^{-5}$] given in every $0.5\log_{10}T$ interval; the blue curves correspond to boxes at lower heights (source A in a3, b3, c3 and source B in d3) and the orange curves to boxes at higher heights (source B in b3, c3 and source C in d3, e3). The animation accompanying this figure shows X-ray sources superimposed upon time-elapsed 131~{\AA} image, temperature map, and EM map corresponding to three different temperature ranges (10.0--12.6, 12.6--17.8, and 17.8--28.2 MK) from about 15:50 to 18:15 UT on 2022 August 28. \label{fig:sc_evo}}
\end{figure*}

\citet{Zhou2022} proposed that a double coronal source is associated with distinct flare-loop systems straddling over the two segments of a $\Gamma$-shaped PIL, but appears to be one above the other in projection.
To test this idea, we examined the magnetic field at the solar limb from two perspectives (Figure \ref{fig:fp_plus_B}). Overall the PIL runs in a north-south direction, but it is not straight either; so it remains possible that hot flaring loops with different orientations are present in the line of sight, leading to the apparent coexistence of plasmas with distinct properties at the loop top, as inferred from the DEM analysis and X-ray spectral fitting.  

In addition, \citet{Kim2014} reported two coronal X-ray sources located at the loop top and slightly above it, respectively, during the decay phase of an M-class flare. They observed plasma flowing from the loop top toward the SAF and inferred upward energy transport, which could be explained by thermal plasma partially escaping magnetic confinement and rising through a ballooning instability \citep{Shibasaki2001} in turbulent plasmas with $\beta\ge 1$. The supra-arcade plasma indeed exhibits turbulent properties \cite[e.g.,][]{Liu&Wang2021,Xie2025}; further, plasma $\beta$ therein is slightly above unity, as inferred from kink waves resulting from the interaction between an SAD and a spike in a companion study on the same event \citep{Pan2025}. However, the upper source in our case is associated with hotter plasma, which suggest downward rather than upward energy transport.

In conclusion, the coronal X-ray emission under investigation exhibits multi-phase evolution and a double coronal source during the flare decay phase, whose energy-height distributions are distinctive from those typically observed during the flare impulsive phase. These results highlight the spatio-temporal complexity of the post-flare current sheet as observed from a side-on perspective, and shed new light on its evolution and energy release process during the flare decay phase.

\begin{acknowledgments}
This work was supported by the Strategic Priority Program of the Chinese Academy of Sciences (XDB0560102), the National Key R\&D Program of China (2022YFF0503002), and the National Natural Science Foundation of China (NSFC; 42274204, 12373064, 42188101, 11925302).
A.M.V. acknowledges the Austrian Science Fund (FWF) projects no. 10.55776/I4555 and 10.55776/PAT7894023. 
Hanya Pan acknowledges the support from China Scholarship Council.

\end{acknowledgments}

\begin{contribution}
Hanya Pan analyzed the data and wrote the first draft. Astrid M. Veronig and Rui Liu devised the research plan and supervised the study. All authors contributed to reviewing and editing the manuscript.

\end{contribution}

\bibliography{draft}{}
\bibliographystyle{aasjournalv7}

%



\end{document}